\documentclass{article}

\usepackage{times}
\usepackage{graphicx} 
\usepackage{subfigure} 

\usepackage{natbib}
\usepackage{array}
\usepackage{algorithm}
\usepackage{algorithmic}

\usepackage{hyperref}

\usepackage[accepted]{icml2014}

\icmltitlerunning{Supervised Convolutional GSN for Protein Secondary Structure Prediction}

\begin{document} 

\twocolumn[
\icmltitle{Deep Supervised and Convolutional Generative Stochastic Network for\\
 Protein Secondary Structure Prediction}

\icmlauthor{Jian Zhou}{jzthree@princeton.edu}
\icmlauthor{Olga G. Troyanskaya}{ogt@cs.princeton.edu}
\icmladdress{Princeton University
           Princeton, NJ 08540 USA}

\icmlkeywords{structured output, generative stochastic network, convolution, secondary structure}

\vskip 0.3in
]

\begin{abstract} 
Predicting protein secondary structure is a fundamental problem in protein structure prediction. Here we present a new supervised generative stochastic network (GSN) based method to predict local secondary structure with deep hierarchical representations. GSN is a recently proposed deep learning technique \cite{Bengio2013b} to globally train deep generative model. We present the supervised extension of GSN, which learns a Markov chain to sample from a conditional distribution, and applied it to protein structure prediction. To scale the model to full-sized, high-dimensional data, like protein sequences with hundreds of amino-acids, we introduce a convolutional architecture, which allows efficient learning across multiple layers of hierarchical representations. Our architecture uniquely focuses on predicting structured low-level labels informed with both low and high-level representations learned by the model. In our application this corresponds to labeling the secondary structure state of each amino-acid residue. We trained and tested the model on separate sets of non-homologous proteins sharing less than $30\%$ sequence identity. Our model achieves $66.4\%$ Q8 accuracy on the CB513 dataset, better than the previously reported best performance $64.9\%$  \citep {Wang2011} for this challenging secondary structure prediction problem.
\end{abstract} 

\section{Introduction}
\label{submission}

Understanding complex dependency between protein sequence and structure is one of the greatest challenges in computational biology \cite{Cheng2008}. Although it is widely accepted that the amino-acid sequence contains sufficient information to determine the three dimensional structure of a protein, it is extremely difficult to predict protein structure based on sequence. Understanding protein structure is critical for analyzing protein function and applications like drug design. Protein secondary structure prediction determines structural states of local segments of amino-acid residues, such as $\alpha$-helix and $\beta$-strand, and it provides important information for further elucidating the three-dimensional structure of protein. Thus it is also being used as input by many other protein sequence and structure analysis algorithms.

Protein secondary structure prediction has been extensively studied with machine learning approaches \cite{Singh2005}.The key challenge in this field is predicting for protein sequences with no close homologs that have known 3D structures. Since the early work by \cite{Qian1988}, neural networks have been widely applied and are core components of many most successful approaches \cite{Rost1993, Jones1999, Baldi1999}.  Most significant improvement to secondary structure prediction has been achieved by leveraging evolutionary information by using sequence profiles from multiple-sequence alignment \cite{Rost1993} or position-specific scoring matrices from PSI-BLAST \cite{Jones1999}. Other developments include better capturing spatial dependencies using bidirectional recurrent neural networks (BRNN) \cite{Baldi1999, Pollastri2002}, probabilistic graphical models \cite{Schmidler2000, Chu2004, van2011hidden}, or combination of neural network and graphical models, like conditional neural fields (CNF) which integrate a windows-based neural network with conditional random field (CRF) \cite{Peng2009,Wang2011}. Secondary structures are commonly classified to 8 states \cite{Kabsch1983} or be further combined into 3 states \cite{Singh2005}. Most secondary structure prediction studies have been focusing on coarse-grained 3-state secondary structure prediction. Achieving fine-grained 8-state secondary secondary prediction, while it reveals more structural details than 3-state predicitons, is a more challenging and less-addressed problem \cite{Pollastri2002, Wang2011}. Here we address the 8-state classification problem for proteins with no close homologs with known structure.

It is widely believed that introducing global information from the whole protein sequence is crucial for further improving prediction performance of local secondary structure labels, since secondary structure formation depends on both local and  long-range interactions. For example, a secondary structure state $\beta$-strand is stabilized by hydrogen bonds formed with other $\beta$-strands that can be far apart from each other in the protein sequence. Many contemporary methods such as BRNN can capture some form of spatial dependency, but they are still limited in capturing complex spatial dependency. To our knowledge none of these methods learns and leverages deep hierarchical representation, which has enabled construction of better intelligent systems interpreting various types of other complex structured natural data, like images, speech, and text. Models with deep representation have been exceptionally successful in capturing complex dependency in these data \cite{Bengio2013a, Salakhutdinov2009}. Learning features automatically can be especially helpful for proteins, as we lack the necessary intuition or knowledge for hand-crafting mid- and high- level features involving multiple amino acids. 

In this work, we introduce a series of techniques to tackle challenges of protein secondary structure prediction with deep learning, and we demonstrate their superior accuracy on 8-state protein secondary structure prediction over previous methods. Although we focus on the secondary structure prediction application here, our methods are general and can be applied to a broad range of problem within and outside of computational biology. First, we extended the recently developed generative stochastic network (GSN) technique to supervised structured output prediction as proposed in \cite{Bengio2013a}. Supervised GSN learns a Markov chain to sample from the distribution of output conditioned on input data, and like GSN it avoided intractable explicit marginalization over hidden variables in learning deep models with hidden layers. The advantage of the supervised GSN over GSN is analogous to the advantage of conditional random field (CRF) over Markov random field (MRF), or discriminative versus generative classifiers, that is, we focus the resources on dependencies that are important for our desired prediction, or the distribution of labels conditional on data. The supervised GSN can be applied to generate structured output of secondary structure states for all amino-acid residues of each protein.  

Another important element of our approach that enables learning hierarchical representation of full-size data, like protein sequences with hundreds of amino-acids, is introducing a convolutional architecture for GSN that  allows for learning multiple layers of hierarchical representations. The convolutional architecture also allows efficient “communication” between high- and low- level features and features at different spatial locations. Convolutional GSN is well suited for making predictions sensitive to low-level or local information, while informed with high-level or global information.

\section{Preliminaries}

\subsection{Generative Stochastic Networks}
\begin{figure*}[ht]
\vskip 0.2in
\begin{center}
\centerline{\includegraphics[width=\textwidth]{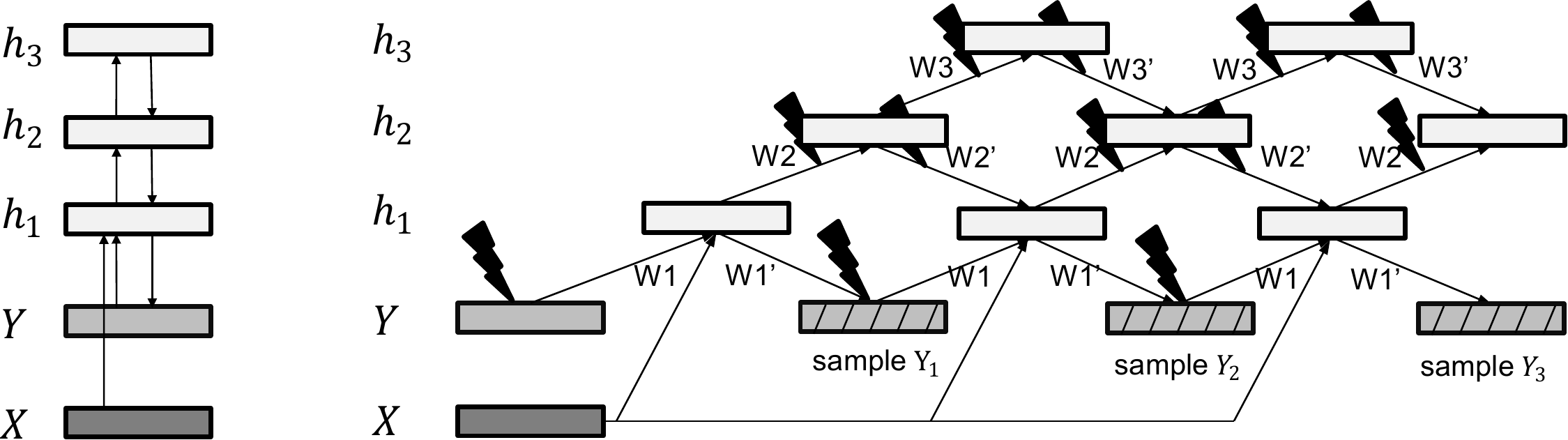}}
\caption{Architecture and computational graph of a supervised convolutional GSN. Connectivity of network is simplified to show only layer-wise connections without details of convolution and pooling. Original and reconstructed labels are corrupted by noise injection which randomly set half of the values to zero (lightning shape), and all input and output from $h_2$ and $h_3$ layers are corrupted by adding Gaussian noise (lightning shape). Each $Y_t$ is obtained by sampling from the $t$-th reconstruction distribution. The log-likelihood of true label $Y$ under each reconstruction distribution is computed and used as the training objective.}
\label{conv_compgraph}
\end{center}
\vskip -0.2in
\end{figure*} 
The generative stochastic network (GSN) is a recently proposed model that utilizes a new unconventional approach to learn a generative model of data distribution without explicitly specifying a probabilistic graphical model, and allows learning deep generative model through global training via back-propagation. Generative stochastic network learns a Markov chain to sample from data distribution $P(X)$. During training, GSN trains a stochastic computational graph to reconstruct the input $X$, which generalizes the generalized denoising auto-encoder \cite{Bengio2013c}. The difference of GSN is that it allows the computational graph to have a latent state, and noise is injected to not just the input, but also to the intermediate computations.  

Compared to other deep generative models like deep Boltzmann machine, a major advantage of GSN is that it avoided intractable inference and explicit marginalization over latent variables, as it does not explicitly learn parameters of a probabilistic graphical model but rather learns to directly sample from the data distribution. GSN also enjoys feasible global training by back-propagating the reconstruction cost.

The following are proven in \cite{Bengio2013c} and \cite{Bengio2013a} as theoretical basis for generalized denoising autoencoder and generative stochastic network:

{\bf Theorom 1.} If $ P_\theta(X|\tilde{X})$ is a consistent estimator of the true conditional distribution $ P(X|\tilde{X})$ ,and $T_n$ defines an irreducible and ergodic Markov chain, then as $n\rightarrow\infty$, the stationary distribution of the generated samples $\pi(X)$ converges to the data generating distribution $P(X)$.

In  \cite{Bengio2013a} this result is extended to introduce a latent state $H_t$ and replace the adding noise and denoising operators with $H_{t+1 }\sim P_{\theta_1}(H|H_t,X_t)$ and $X_{t+1 }\sim P_{\theta_2}(X|H_{t+1})$ respectively.

{\bf Theorem 2.}  Let training data $X \sim P(X)$ and independent noise $Z \sim P(Z)$ and introduce a sequence of latent variables $H$ defined iteratively through a function $f$ with $H_t=f_{\theta_1}(X_{t-1},Z_{t-1},H_{t-1})$ for a given sequence of $X_t$'s. Consider a model $ P_{\theta_2} (X | f_{\theta_1} (X, Z_{t-1},H_{t-1}))$ trained (over both $\theta_1$ and $\theta_2$) so that for a given $\theta_1$, it is a consistent estimator of the true $P(X|H)$. Consider the Markov chain defined above and assume that it converges to a stationary distribution $\pi_n$ over the $X$ and $H$ and with marginal $\pi_n(X)$, even in the limit as number of training examples $n\rightarrow\infty$. Then $\pi_n (X)\rightarrow P(X)$ as $n\rightarrow\infty$.

It is suggested in \cite{Bengio2013a} that the GSN can be extended to tackle structured output problems. As focusing on conditional distribution while sharing the advantages of GSN seems to provide an attractive new approach of leveraging deep representations for structured output prediction. We explored this idea of supervised generative stochastic network in the following.

\subsection{Supervised Generative Stochastic Network}

We phrase the supervised generative stochastic network as follows. We use X to denote input random variable and Y to denote output random variable. Supervised GSN aims to capture conditional dependency structure of Y given X. Analogous to GSN, it learns a Markov chain that samples from $P(Y|X)$. 

Let $C(\tilde{Y}|Y)$ be a corruption process that transforms $Y$ into a random variable $\tilde{Y}$. Let $P_\theta (Y|\tilde{Y},X)$ be a denoising auto-encoder that computes probability of $Y$ given $\tilde{Y}$ and $X$. Then we have the following.

{\bf Corollary 1.} If $P_\theta (Y|\tilde{Y},X)$ is a consistent estimator of the true conditional distribution $P(Y|\tilde{Y},X)$, and $T_n$ defines an irreducible and ergodic Markov chain, then as $n\rightarrow\infty$, the stationary distribution of the generated samples $\pi(Y|X)$ converges to the data generating distribution $P(Y|X)$.

{\bf Proof.} Let $ P(X')= P(Y | X)$; $C(\tilde{X'}| X') = C(\tilde{Y}|Y) $;  $P_\theta (X'| \tilde{X'} )=P_θ (Y| \tilde{Y},X)$; Then this corollary is equivalent to theorem 1.

During training, we learn a computational graph that estimates $P_\theta (Y|\tilde{Y},X)$. Therefore we provide both $\tilde{Y}$ and $\tilde{X}$ as input and the training procedure should minimize reconstruction cost. The reconstruction cost should be selected to be interpretable as (regularized) log likelihood of uncorrupted labels given the reconstructon distribution. 

To generalize the supervised training to generative stochastic networks, similar to Corollary 1 we consider independent noise $Z\sim P(Z)$.  Corruption process is now $H_{t}=f_{\theta_1} (Y_{t-1},Z_{t-1},H_{t-1},X)$, where $f_{\theta_1}$ is an arbitrary differentiable function. The reconstruction function $P_{\theta_2} (Y|f_{\theta_1} (Y_{t-1},Z_{t-1},H_{t-1},X))$ is trained by regularized conditional maximum likelihood with examples of $(Y,Z,X)$ triplets. 

{\bf Corollary 2.} Assume that the trained model $P_{\theta_2} (Y|f_{\theta_1} (Y_{t-1},Z_{t-1},H_{t-1},X))$ is a consistent estimator of the true $ P(Y|H)$, and the Markov chain $Y\sim P_{\theta_2} (Y|f_{\theta_1} (Y_{t-1},Z_{t-1},H_{t-1},X))$ converges to a stationary distribution $\pi$. Then we have $\pi(Y|X)=P(Y | X)$ following Theorem 2.

To make a prediction from a trained supervised GSN model, we can sample from $P_{\theta_1,\theta_2}(Y| X)$ by initializing from arbitrary Y and then running sampling via Markov chain $Y\sim P_{\theta_2} (Y|f_{\theta_1} (Y_{t-1},Z_{t-1},H_{t-1},X))$. 

Supervised GSN allows flexible network architecture and noise forms just as GSN.  It enjoys the same advantage of avoiding marginalization over hidden variables in both training and inference. Both Supervised GSN and GSN can benefit from converting a difficult learning task of capturing $P(X)$ or $P(Y|X)$ into easier tasks of learning reconstruction distributions.

\section{Algorithms}

\subsection{Convolutional Generative Stochastic Network}

\begin{figure}[ht]
\vskip 0.2in
\begin{center}
\centerline{\includegraphics[width=\columnwidth]{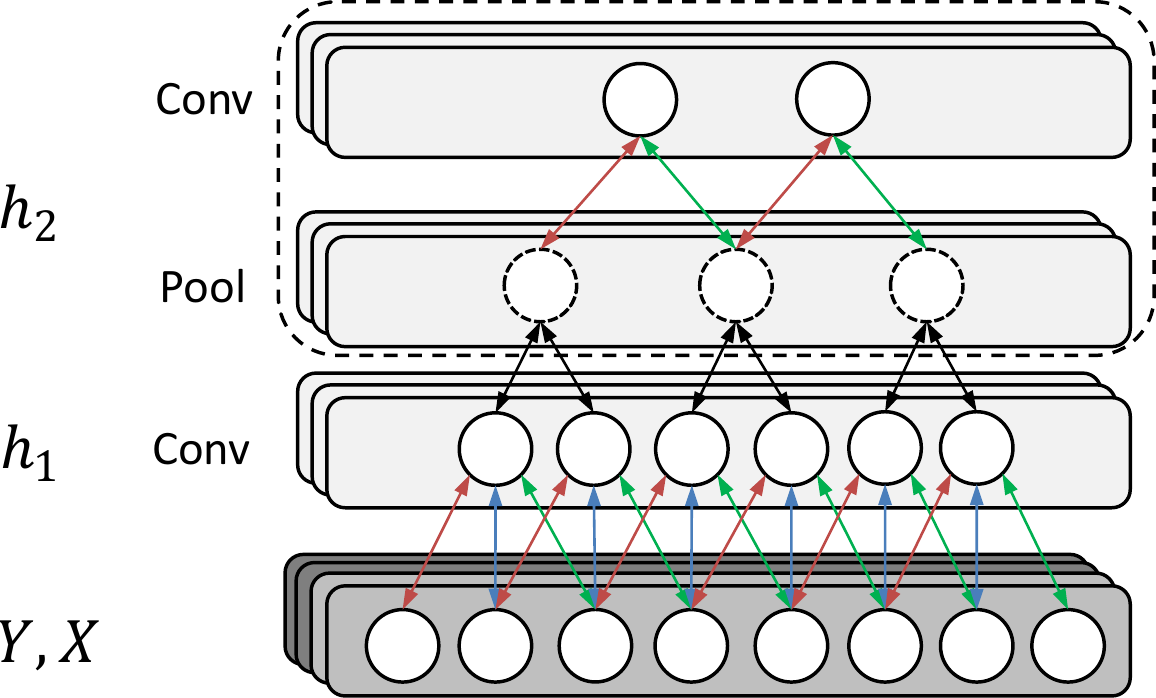}}
\caption{Show the spatial connectivity architecture of a convolutional GSN architecture with 2 convolutional GSN layers. Each convolution layer (conv) feature is connected to all feature maps of the lower layer, while pooling layer (pool) features connect only to features in the corresponding feature map. In the computational graph of convolutional GSN (Figure~\ref{conv_compgraph}), a convolutional GSN layer may contain a single convolutional layer ($h_1$) or both a pooling and a convolutional layer ($h_2$), and pooling layers are considered as intermediate computations for calculating input of convolutional layers.}

\label{conv_connection}
\end{center}
\vskip -0.2in
\end{figure}

The original  ‘DBM-like’ GSN architecture proposed in \cite{Bengio2013a} does not leverage the spatial structure of the data, therefore each feature is connected to data for every location, making it hard to scale to real-sized data. Therefore we introduce convolutional GSN (Figure~\ref{conv_compgraph}, \ref{conv_connection}) to learn a hierarchical representation with gradually more global representation in higher levels. Given the sequential nature of protein data, we use 1D convolution here, but the architecture can be easily generalized to higher-dimensional convolution. This architecture is applicable to both GSN and supervised GSN.

The simplest convolutional GSN includes only an input layer and a convolution layer. Assume the input layer consists of  $N_v\times C_v$ visible units, where $N_v$ is the size of input and $C_v$ is number of input channels, and the convolutional layer includes $K$ feature maps, and each map is associated with a size $N_w\times C_v$ filter, where $N_w$ is the convolutional window size. The convolution layer thus consists of  $(N_v-N_w+1)×K$ hidden units each connects with $N_w\times C_v$ visible units. The filter weights are shared across the map.  Each map also has a hidden unit bias $b$ and a visible unit bias $b'$.  We use $*$ to denote 1D convolution, then the upward and downward pass between visible layer and convolutional layer is computed as:\[y^k=\tilde{\sigma}(\tilde{x}*W^k+b^k,z)\] \[x'=\sigma(\sum_{k}y^k*W'^k+b'^k)\]

Noise is injected through an independent random variable $z\sim P(Z)$  with the noisy activation function $\tilde{\sigma}$. In this work we used the same noisy activation function as used in \cite{Bengio2013a}: \[\tilde{\sigma}(x,z)= a_{post} z+\textrm{tanh}⁡(x+a_{pre} z)\] Gradient over stochastic neurons is calculated by only considering differentiable part, or ``straight-through estimator'', as in \cite{Bengio2013b}. 

Convolutional layers can be stacked to create a deep representation. To allow learning high-level representations, it is usually desirable to reduce the size of feature maps in upper layers by pooling. Therefore we applied mean-pooling that pools several adjacent features in a feature map and output their average value. An alternative to mean-pooling is max-pooling, which outputs max rather than average value; because the max operation is not reversible, we can do the downward pass as if the upward pass used mean pooling. Pooling also effectively hard-codes translational invariance in higher level features. 

In the computational graph of convolutional GSN, we use layer-wise sampling similar to the “deep Boltzmann machine-like” network architecture proposed in \cite{Bengio2013a}. As pooling layer contains redundant information with the lower layer, it is considered together with its upper-level convolutional layer as a convolutional GSN layer. Intuitively in the layer-wise sampling, pooling layers can be considered as only providing intermediate computation for input of convolutional layers, and convolutional layers activations are stored and used for the next sampling step.

Note that the recurrent structure of its computational graph makes convolutional GSN significantly different from the feed-forward convolutional networks. In convolutional GSN, the representations from different layers and different spatial positions can “communicate” with each other through bidirectional connections. With a sufficient number of alternating upward and downward passes, information from any position of the input/feature map can be propagated to any other positions through multiple different paths (the number of passes required depends directly on the architecture; if the top layer feature is connect to all of the input positions, then $2N_{layer}$ rounds is enough). 

While the depth of representation can be considered as the number of layers in convolutional GSN, the depth of computational graph grows as the number of `sampling' iterations, which is also called number of walkbacks \cite{Bengio2013a},  increases, so the computational graph can be arbitrarily deep with replicated parameters. 

When using this convolutional GSN architecture for supervised training like protein secondary structure prediction (Figure~\ref{conv_compgraph}), we will consider two types of input channels, feature channels (X) and label channels (Y). In the computational graph of supervised convolutional GSN  (Figure~\ref{conv_compgraph}), we only corrupt and reconstruct label channels, and feature channels are not corrupted and provided as part of input to compute the activations of the first hidden layer.

\section{Experimental Results} 
 
\subsection{Features and dataset}

Our goal is to correctly predict secondary structure labels for each amino-acid of any given protein sequence. We focused on 8-state secondary structure prediction as this is a more challenging problem than 3-state prediction and reveals more structural information. We used multi-task learning to simultaneously predict both secondary structure and amino-acid solvent accessibility, because learning to predict other structural properties with the same network allows sharing feature representations and may thus improve feature learning.

Evolutionary information as position-specific scoring matrix (PSSM) is considered the most informative feature for predicting secondary structure from previous research \cite{Jones1999}.  To generate PSSM, which are $ n \times b$ matrices, where n is protein length and b is the number of amino-acid types, we ran PSI-BLAST against UniRef90 database with inclusion threshold 0.001 and 3 iterations. We used pfilt program from PSI-PRED package to pre-filter the database for removing low information content and coiled-coil like regions \cite{Jones1999}. To use PSSM as the input for the supervised GSN model, we transform the PSSM scores to 0-1 range by the sigmoid function. We also provide the original protein sequence for amino-acid residues encoded by $ n \times b$ binary matrices as input features. Two other input features encode start and end positions of the protein sequence. While other features could be considered to further improve performance, we focus here on evaluating the model architecture.

We used a large non-homologous sequence and structure dataset with $6128$ proteins (after all filtering), and divided it randomly into training (5600), validation (256), and testing (272) sets. This dataset is produced with PISCES Cull PDB server \cite{Wang2003} which is commonly used for evaluating structure prediction algorithms. We retrieved a subset of solved protein structures with better than 2.5\AA\   resolution while sharing less than 30\% identity, which is the same set up as used in \cite{Wang2011}. We also removed protein chains with less than 50 or more than 700 amino acids or discontinuous chains.

8-states secondary structure labels and solvent accessibility score were inferred from the 3D PDB structure by the DSSP program \cite{Kabsch1983}.  We discretized solvent accessibility scores to absolute solvent accessibility and relative solvent accessibility following \cite{Qi2012}.  

The resulting training data including both feature and labels has 57 channels (22 for PSSM, 22 for sequence, 2 for terminals, 8 for secondary structure labels, 2 for solvent accessibility labels), and the overall channel size is 700. The 700 amino-acids length cutoff was chosen to provide a good balance between efficiency and coverage as the majority of protein chains are shorter than 700AA. Proteins shorter than 700AA were padded with all-zero features.

We evaluated performance of secondary structure prediction on the test dataset containing 272 proteins. We also performed a seperate evaluation on CB513 dataset while training on Cull PDB dataset further filtered to remove sequences with $>25\%$ identity with the CB513 dataset. Secondary structure prediction performance was measured by Q8 accuracy, or the proportion of amino-acids being correctly labeled. We ran 5 parallel Markov chains each generating 5000 consecutive reconstruction distributions from the model and evaluated performance by averaging the last 500 reconstruction distributions from each Markov chain. 

\subsection{Training setup}

To inject noise to labels, the input labels were corrupted by randomly setting half of the values to zero. Gaussian noise with standard deviation 2 was added pre- and post- activation on all except the first convolutional GSN layer. Each $y_t$ was obtained by sampling from the reconstructed distribution. We consider reconstructed distribution as multinomial distribution for secondary structure and binomial distribution for solvent accessibility. Walkback number is set to $12$ to ensure flow of information across the whole network. 

Motivated to better learn the reconstruction distribution when the chain is initilized by an arbitrary label $Y'$ far away from where the probability density of true $P(Y|X)$ lies, which is important for supervised prediction tasks, during training for half of the training samples we initalized Y by the same arbitrary initiation as used for prediction rather than its true value. We call this trick "kick-start". 

Tanh activation function is used for all cases except the visible layer, which uses sigmoid activation function for reconstruction. All layers use untied weights, so W and W' do not equal.  We regularize the network by constraining the norm of each weight vector.

Parameters for all layers were trained globally by back-propagating the reconstruction error. Mini-batch stochastic gradient descent with learning rate 0.25 and momentum 0.5 was used for training model parameters. All models for comparisons were trained for 300 epochs. The model is implemented based on Theano and Pylearn2 libraries and trained on Tesla K20m GPU.

\subsection{Performance}

\begin{table}[t]
\caption{Comparison of different architectures.}
\label{sample-table1}
\vskip 0.15in
\begin{center}
\begin{small}

\begin{tabular}{lc >{\centering\arraybackslash}p{0.8in} }
\hline
\abovespace\belowspace
Model & Q8 Accuracy  & Segment Overlap Score \\
\hline
\abovespace
SC-GSN-3layer  &  ${\bf 0.721}\pm0.006$ &  ${\bf0.695}\pm0.006$ \\
SC-GSN-3layer  & $0.711\pm0.006$   & $0.689\pm0.007$\\ (without "kick-start")\\ 
SC-GSN-3layer  & $0.687\pm0.006$   & $0.665\pm0.007$\\ (without Gaussian noise)\\ 
SC-GSN-2layer &  $0.720\pm0.006$ &  $0.695\pm0.007$ \\
SC-GSN-2layer  & $0.713\pm0.006$   & $0.688\pm0.007$\\ (without "kick-start")\\ 
SC-GSN-2layer  & $0.698\pm0.007$   & $0.681\pm0.007$\\ (without Gaussian noise)\\ 
\belowspace
SC-GSN-1layer   &  $0.714\pm0.006$  &  $0.689\pm0.007 $\\

\hline
\end{tabular}

\end{small}
\end{center}
\vskip -0.1in
\end{table}

\begin{table}[t]
\caption{Performance on public benchmark dataset CB513}
\label{sample-table1}
\begin{center}
\begin{small}

\begin{tabular}{lcccr}
\hline
\abovespace\belowspace
Model & Q8 Accuracy   \\
\hline
\abovespace
SC-GSN-3layer  &  ${\bf 0.664}\pm0.005$ \\
\belowspace
Previous state-of-the-art \citep {Wang2011}  &  $0.649\pm0.003$ \\

\hline
\end{tabular}

\end{small}
\end{center}
\vskip -0.2in
\end{table}

\begin{table}[t]
\caption{Classification accuracies for individual secondary structure states.\\  \\
Individual secondary structure state prediction sensitivities, precisions, and frequecies calculated across amino acid residues of all proteins in our Cull PDB test set are shown below. State I cannot be evaluated as it is too rare to even appear in the test set. }
\label{sample-table2}

\begin{center}
\begin{small}

\begin{tabular}{lcccr}
\hline
\abovespace\belowspace
Sec. & Sensitivity & Precision  & Frequency &Description  \\
\hline
\abovespace
H &    0.935 & 0.828 & 0.354& $\alpha$-helix\\
E     &   0.823 & 0.748 & 0.218& $\beta$-strand \\
L  &  0.633  & 0.541  &0.186&  loop or irregular \\
T      &   0.506 & 0.548  & 0.111 &  $\beta$-turn \\
S   &    0.159 & 0.423  & 0.079& bend \\
G    &  0.133 & 0.496 & 0.041& $3_{10}$-helix   \\
B      &  0.001 & 0.5 &  0.011  & $\beta$-bridge \\
\belowspace
I    &    -   &  - & 0 & $\pi$-helix\\

\hline

\end{tabular}

\end{small}
\end{center}

\vskip -0.1in
\end{table}

\begin{figure*}[ht]
\vskip 0.2in
\begin{center}
\centerline{\includegraphics[width=0.7\textwidth]{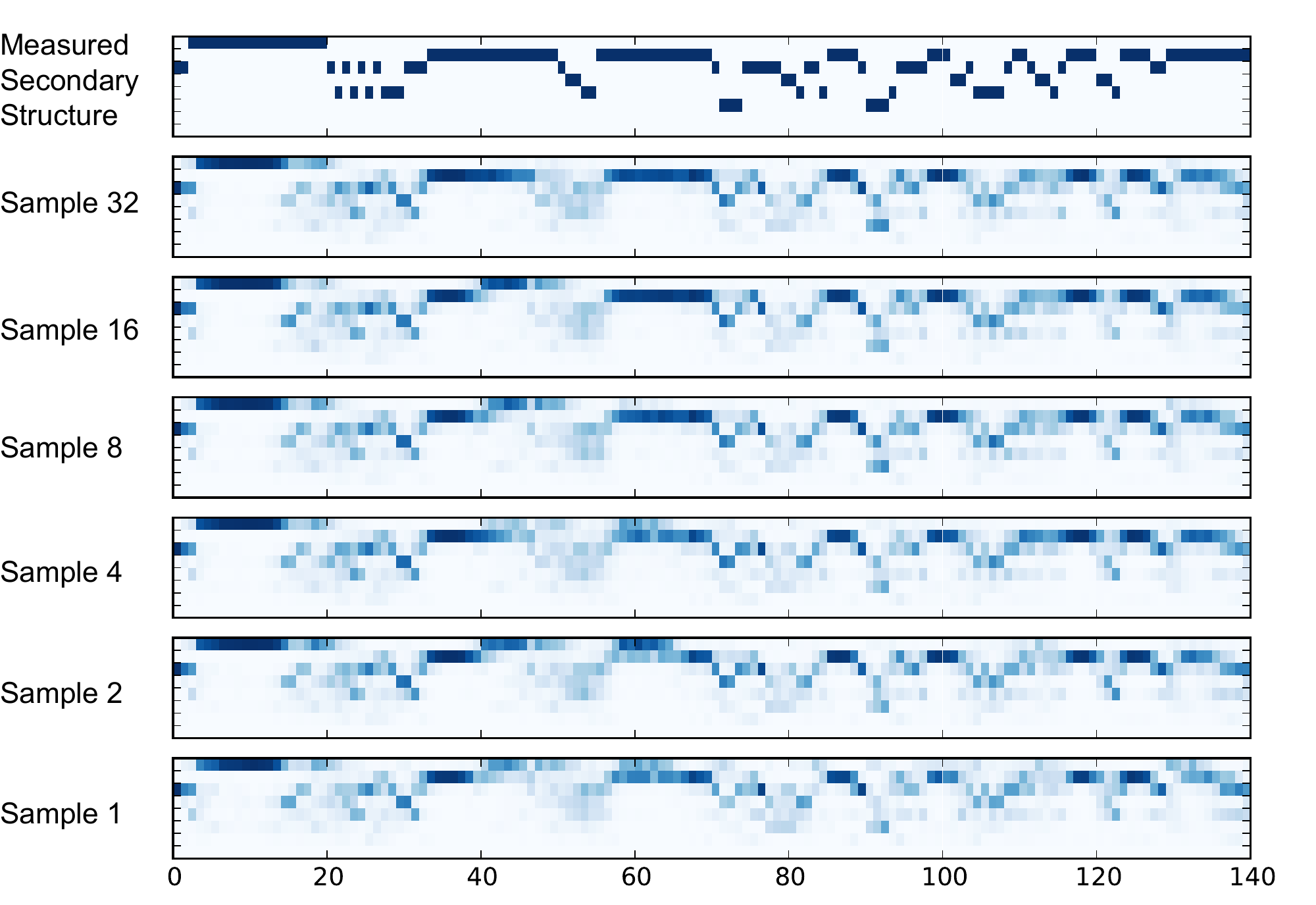}}
\caption{Secondary structure prediction example for a test set protein from a trained model. Each row in a sub-panel represent a secondary structure state, which are H, E, L, T, S, G, B, and I from top to bottom. See Table~\ref{sample-table2} for descriptions of secondary structure states. Each column corresponds to one amino-acid residue in this protein sequence, and X-axis  shows indices of amino acid residues in this protein. Dark versus light color indicates strong versus weak confidence of prediction.}
\label{example_1}
\end{center}
\vskip -0.2in
\end{figure*}

\begin{figure}[ht]
\vskip 0.2in
\begin{center}
\centerline{\includegraphics[width=\columnwidth]{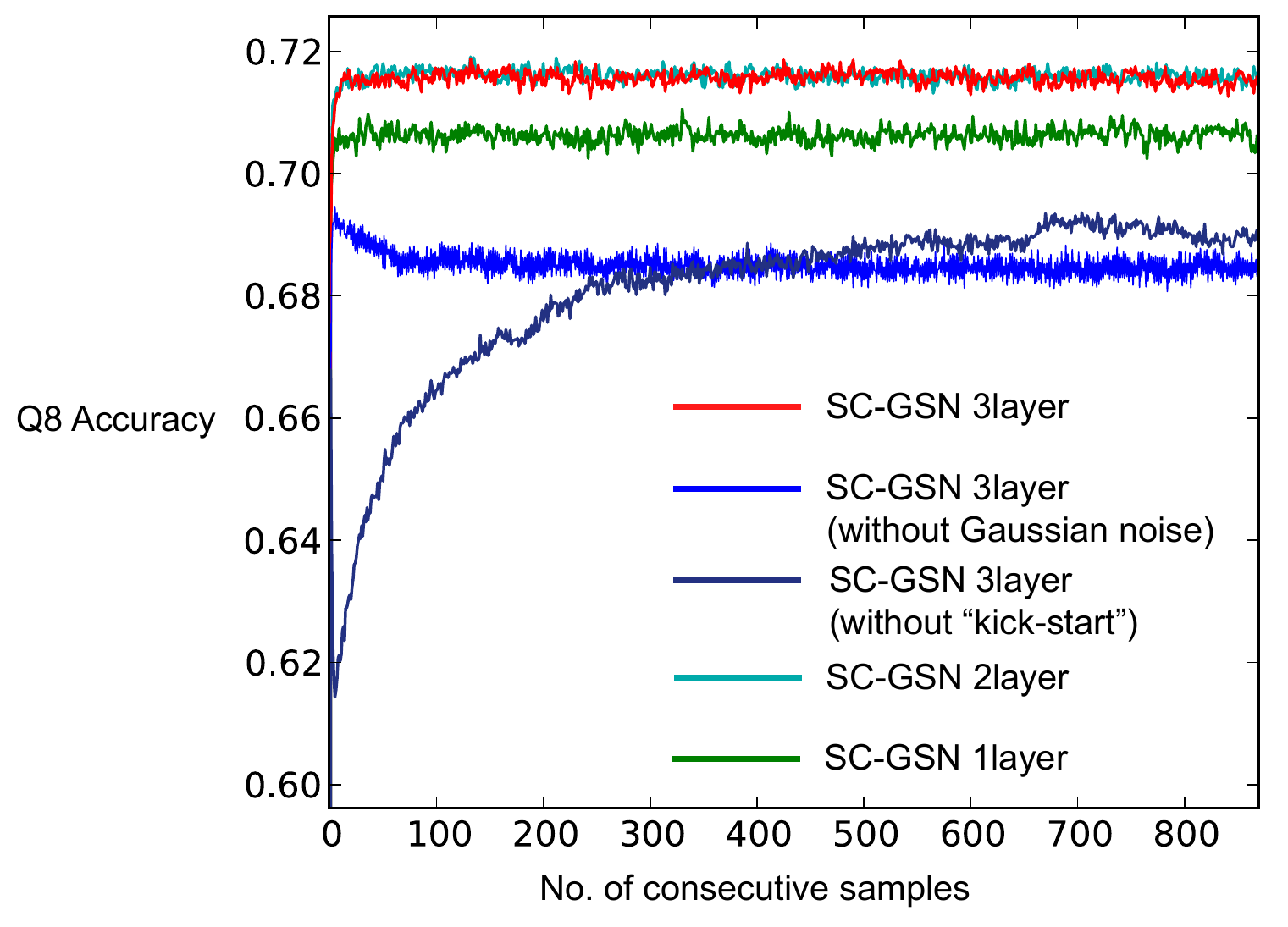}}
\caption{ Prediction performances of individual consecutive samples from trained models with different architectures. Q8 accuracies are calculated for our Cull PDB test set proteins at each sampling iteration for each model. Average perfomances across 272 test proteins are shown. }
\label{example_2}
\end{center}
\vskip -0.2in
\end{figure} 

We achieved $72.1\pm0.6\%$ Q8 accuracy on our Cull PDB test set sequences (Table~\ref{sample-table1}). The best single model we tested features a 3-layer convolutional structure. We denote our structure as $\{80\times conv5\}-\{pool5-80\times conv5\}-\{ pool5-80\times conv4\}$. `conv' or `pool' denotes whether the layer type is convolutional or pooling layer, and the number that follows layer type denotes the size of filter or pooling window. Mean pooling is used for all pooling layers in this architecture. We used 80 channels in all convolutional layers.  


Figure~\ref{example_1} shows an example of consecutive samples generated from the trained model conditional on input features from a test set protein. We observe that as the chain runs the samples get gradually closer to the experimentally measured secondary structure, and some incorrect predictions are fixed in later steps of sampling. The later samples also seems to have more spatially consistent organization. 

Prediction performance for individual secondary structure states are shown in (Table 2). We achieved high accuracy for four major states H, E, L and T. Prediction for less frequent states G and S are known to be difficult, because they have a limited number of training standard leading to significantly unbalanced labels. We did not specifically try to address the unbalanced label problem here. More efforts should be made in future to better identify these states. State I is extremely rare, so we did not even encounter one in the test set, thus it potentially limits improvement on its prediction. 

We also perfomed validation on a public standard benchmark dataset CB513. For this validation we trained a model aftering filtering the Cull PDB dataset to remove sequences with homology with CB513 sequences (i.e. $>25\%$ identity). Our model achieves Q8 accuracy of 0.664, outperforming previous state-of-the-art result by CNF/Raptor-SS8 (0.649) \cite {Wang2011}.

\subsection{Analysis of architecture}

To discover important factors for the success of our architecture, we experimented with alternative model architectures with varying number of layers, and tried to remove "kick-start" training or Gaussian noise injection (Table~\ref{sample-table1}, Figure~\ref{example_2}).

For model architectures with different number of layers, we removed the top 1 and 2  convolutional GSN layers from the original best 3-layer architecture. The 2-layer model outperforms the 1-layer model, and the 3-layer model has slightly better performance than the 2-layer model.

Not using "kick-start" during training dramatically decreases the speed of convergence to optimal performance during sampling and the prediction accuracy. This is likely because "kick-start" model learned better reconstruction distribution when chain is initalized from an arbitrary state. 

Noise injected in computation of the upper layers seems to be necessary for optimal performance for deep models. We also note that lower reconstruction error on validation data during training does not always correspond to better prediction performance.  For our 3-layer model with no Gaussian noise injected to intermediate computations, although the validation reconstruction error goes lower than for the model with noise, the samples from the model with noise give much better predictions (0.721 vs. 0.687).




\section{Conclusions} 
To apply deep representations to protein secondary structure prediction, we implemented a supervised generative stochastic network (GSN) and introduced a convolutional architecture to allow it to learn hierarchical representation on full-sized data.  While we demonstrated its success on secondary structure prediction, such architecture can be potentially applied to other protein structure prediction tasks.  Our experiments suggest supervised generative stochastic network to be an effective algorithm for structured prediction, extending the success of generative stochastic network in capturing complex dependency in the data. 

The combination of convolutional and supervised generative stochastic network we applied is well suited for low-level structured prediction that is sensitive to local information, while being informed of high-level and distant features. Thus, such an architecture may also be applicable to a wide range of structured prediction problems outside bioinformatics such as scene parsing and image segmentation.

For further development of the protein sequence and structure modeling, one limitation of the current architecture is that the convolutional structure is hard-coded, thus it may not be optimal to capture the spatial organization of protein sequence in some cases, especially for structures formed by long-range interactions. To better model the long-range interactions in a protein, adaptive and dynamic architecture that changes the connectivity adaptively based on input, analogous to unfolding recursive auto-encoder \cite{Socher2011}, may further improve the quality of representation in future.

\section*{Acknowledgments} 
We would like to thank Robert E. Schapire for helpful discussions. We also acknowledge the TIGRESS high performance computer center at Princeton University for computational resource support. This work was supported by National Science Foundation (NSF) CAREER award (DBI-0546275), National Institutes of Health (NIH) (R01 GM071966, R01 HG005998 and T32 HG003284) and National Institute of General Medical Sciences (NIGMS) Center of Excellence (P50 GM071508). O.G.T. is a Senior Fellow of the Canadian Institute for Advanced Research.


\bibliography{ref2}
\bibliographystyle{icml2014}

\end{document}